\documentclass[aps,prl,twocolumn,groupedaddress,showpacs]{revtex4}

\usepackage{graphicx}
\begin{document}

%\preprint{}

\title{Hexacyanometalate Molecular Chemistry : Trinuclear CrNi$_2$ 
Complexes;  
MicroSQUID Magnetisation Studies of Intermolecular Interactions}



\author{Raluca Tiron$^a$, Wolfgang Wernsdorfer$^a$, Fabien Tuyeras$^b$, Ariane Scuiller$^b$,
ValŽrie Marvaud$^b$, Michel Verdaguer$^b$}

%\thanks{}
\affiliation{
$^a$Lab. L. N\'eel, associ\'e \`a l'UJF, CNRS, BP 166, 38042 Grenoble Cedex 9,France\\
$^b$LCIMM, CNRS UMR 7071, BP 42, 
75252 Paris Cedex 05, France\\
}


\date{\today}

\begin{abstract}
Three different CrNi$_2$ complexes were synthesized. They differ from each other by
the nature of the terminal ligand and of the counter anion : 
[Cr(CN)$_4$[CN-Ni(tetren)]$_2$]Cl, that crystallizes in two different crystallographic
systems and [Cr(CN)$_4$[CN-Ni(dienpy$_2$)]$_2$](ClO$_4$). The ground state spin value is 7/2 
for the three systems (ferromagnetic interaction between chromium(III) and
nickel(II) ions). The magnetisation of the three CrNi$_2$ complexes was measured 
using an array of micro-SQUIDs in a temperature range between 0.04 K and 7 K,
under a magnetic field up to ±1 T. The three samples present a three dimensional
magnetic ordering. The correlation between the structure and the intermolecular
coupling is analysed in terms of steric hindrances of the terminal ligand, 
orientation of the molecules in the unit cell (canted structure) and crystal 
symmetry.	
\end{abstract}

%\keywords{}

\maketitle

{\bf 1. Introduction}

\indent The search for new polynuclear molecules displaying high spin ground state
and anisotropy raises the interest of synthetic chemists 
~\cite{1,2,3,4,5,6} and physicists ~\cite{7,8,9} involved in the field of 
molecular magnetism	~\cite{10,11}. 
These anisotropic high spin molecules exhibit original magnetic properties 
such as single molecule magnet behaviour ~\cite{12,13,14,15,16,17} (long
relaxation time for the magnetisation below a so-called blocking temperature,
$T_B$) or magnetic quantum tunnelling effect ~\cite{18,19}. Our research approach
is devoted to the synthesis of such  complexes in order to get single
molecule.
%\begin{figure}
%\begin{center}
%\includegraphics[width=.42\textwidth]{fig_2a.eps}
%\includegraphics[width=.42\textwidth]{fig_2b.eps}
%\includegraphics[width=.42\textwidth]{fig_2c.eps}
%\caption{Projection of the unit cells of 1 (along the b axis) , 
%of 1$^*$ (along the a axis) and of 3 (along the b axis)}
%\label{fig 2}
%\end{center}
%\end{figure}

\indent In order to understand the single molecule magnet (SMM) behaviour, it is 
important to remind that a phenomenological Hamiltonian correlated to a spin
state diagram might describe all polynuclear complexes. In presence of a 
uniaxial anisotropy, a zero-field splitting is observed which splits the ground
state in 2S + 1 levels. The activation energy of the anisotropy barrier, $E_a$,
between the spin states $M_S$ = ± S, is a direct function of D (uniaxial anisotropy)
and S (ground state spin value), $E_a$ = $- DS^2$. To obtain a single molecule magnet
i.e. to avoid a rapid relaxation process, the spin as well as the absolute value
of the anisotropy have to be as high as possible and D has to be negative so
that the state of highest $M_S$ lies below in energy. The exchange coupling J 
between two spin carriers has to be important to well separate the ground state
from the first excited states. The intermolecular interaction, 
$J^{'}_{inter}$, plays 
also an essential role: it has to be negligible, necessary condition to avoid a
three-dimensional magnetic ordering and to observe the properties of an isolated
nanoscale object. Very few examples of clusters responding to all these criteria
have been described in the literature, among them Mn$_{12}$ 
~\cite{14,15,20}, Fe$_8$ ~\cite{21} and 
Mn$_4$ ~\cite{9} are reference compounds.

\begin{figure}
\begin{center}
\includegraphics[width=.42\textwidth]{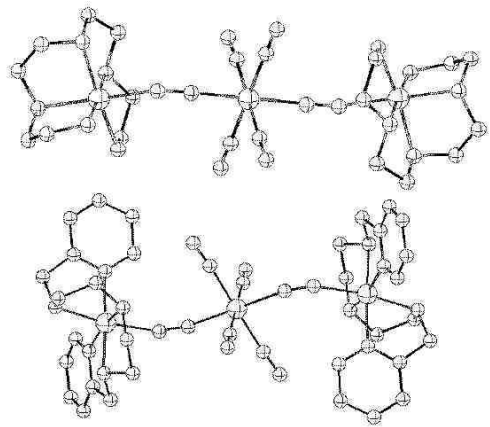}
\caption{X-ray crystal structure of [Cr(CN)$_4$[CN-Ni(tetren)]$_2$]$^+$ 
(up) and 
[Cr(CN)$_4$[CN-Ni(dienpy$_2$)]$_2$]$^+$ (down) entities in 1 and 3.}
\label{fig 1}
\end{center}
\end{figure}

To obtain anisotropic high spin molecules, we developed a step by step synthetic
strategy based on the coupling reaction of polycyanometalate precursors 
(viewed as Lewis bases) and mononuclear complexe
of transition metal ions
with a polydentate ligand, leaving a unique accessible position only
(as Lewis acids) ~\cite{22,23,24}. Starting with hexacyanometalate precursors,
this strategy allows to obtain : i) isotropic high spin molecules such as 
CrCu$_6$ (S = 9/2), CrNi$_6$ (S = 15/2) and CrMn$_6$ (S = 27/2) ~\cite{24};
ii) complexes with various structural anisotropy but low ground state spin,
for instance CoNi (S = 1), CoNi$_3$ (S = 1) and CoNi$_5$ (S = 1) or even singlets
CoNi$_2$ (S = 0) and CoNi$_6$ (S = 0) ~\cite{25}; iii) anisotropic high spin
molecules like CrNi (S = 5/2), CrNi$_2$ (S = 7/2), CrNi$_3$ (S = 9/2) and CrNi$_5$ 
(S = 13/2) ~\cite{25,26,27}.
In the present paper, different CrNi$_2$ complexes
([Cr(CN)$_4$[CN-Ni(L)]$_2$], L
being a polydentate ligand) are investigated below 2K in order to search 
a potential single molecule magnet behaviour or to evaluate the intermolecular
magnetic exchange interaction. After presenting briefly the structural 
parameters of the various complexes, we discuss the low temperature magnetic 
properties on oriented single crystals and evaluate the main parameters 
involved in the intermolecular exchange interaction which induces the 
three-dimensional ordering at very low temperature
(See also Ref.~\cite{28,29,30}). \\

{\bf 2.  CrNi$_2$ complexes, presentation and structure}

\indent Five different [Cr(CN)$_4$[CN-Ni(L)]$_2$]$^+$ complexes,
abbreviated as CrNi$_2$ in the following, have been isolated and characterised,
including by single crystal X-ray diffraction. The compounds differ
from each other by the nature of the polydentate ligand, L, 
[tetraethylenepentamine (tetren), tris(2-aminoethyl)-N,N,N$^{'}$-ethyldiamine (trenen)
and bis(2-pyridylmethyl) - N, N$^{'}$ - diethylenetriamine (dienpy$_2$)] or by the nature
of the counter anions (tetrafluoroborate, chloride or perchlorate). 
The magnetic studies (4-300K) indicate in all cases a ferromagnetic 
intramolecular interaction between the chromium and the nickel spin carriers,
as predicted for orthogonal magnetic orbitals and a S = 7/2 value for 
the ground state. The intramolecular exchange coupling value, J$_{CrNi}$,
varies from + 4.5 cm$^{-1}$ to + 10 cm$^{-1}$ according to the distortion of the
Cr-CN-Ni unit and especially to the change in the C - N - Ni angle. The uniaxial 
anisotropy, D, computed from fitting of the susceptibility data, 
is around - 0.3 cm$^{-1}$. High-field EPR experiments are in progress in order to 
determine precisely the anisotropy factor. The synthesis, the characterisation,
the magnetic properties and magneto-structural correlation studies of these 
species are presented elsewhere ~\cite{24}.

\begin{figure}
\begin{center}
\includegraphics[width=.4\textwidth]{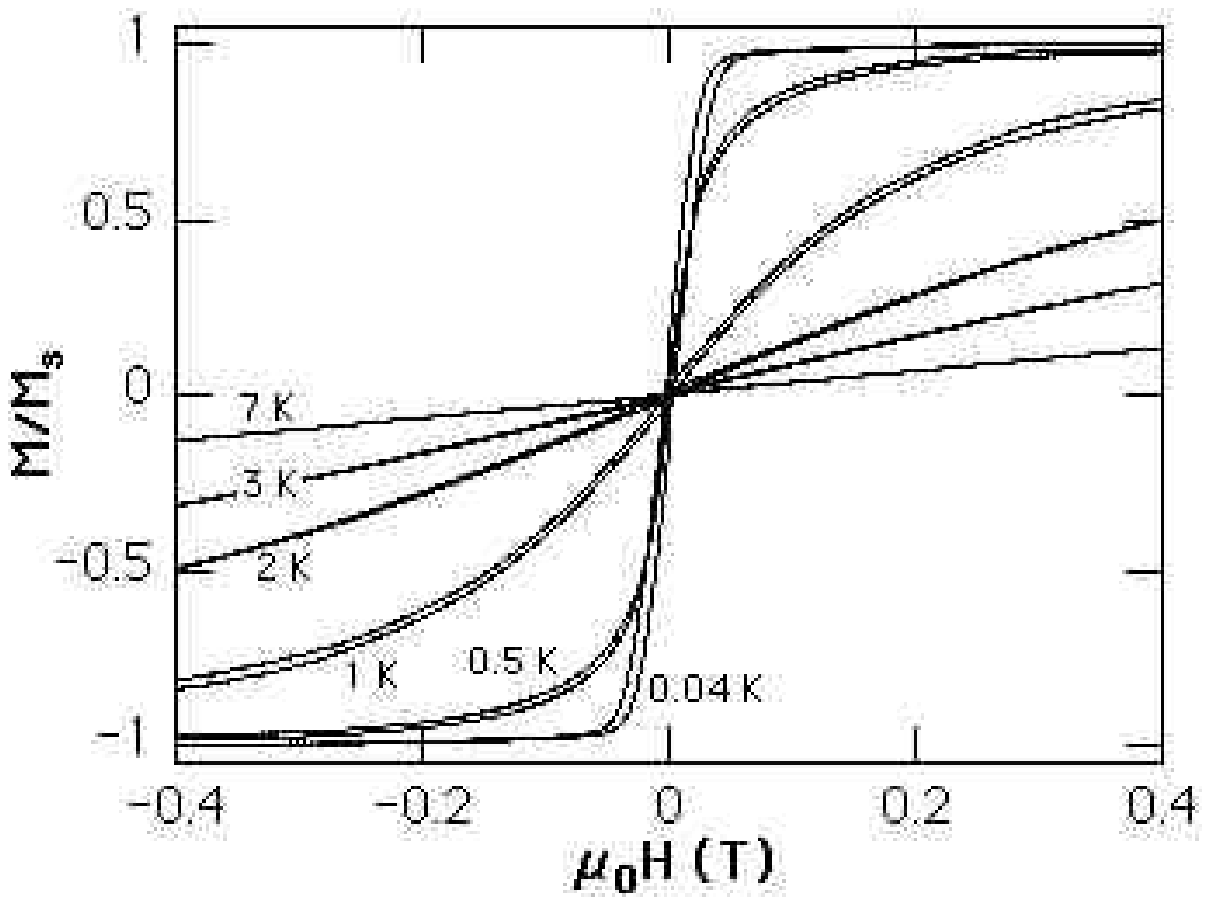}
\includegraphics[width=.4\textwidth]{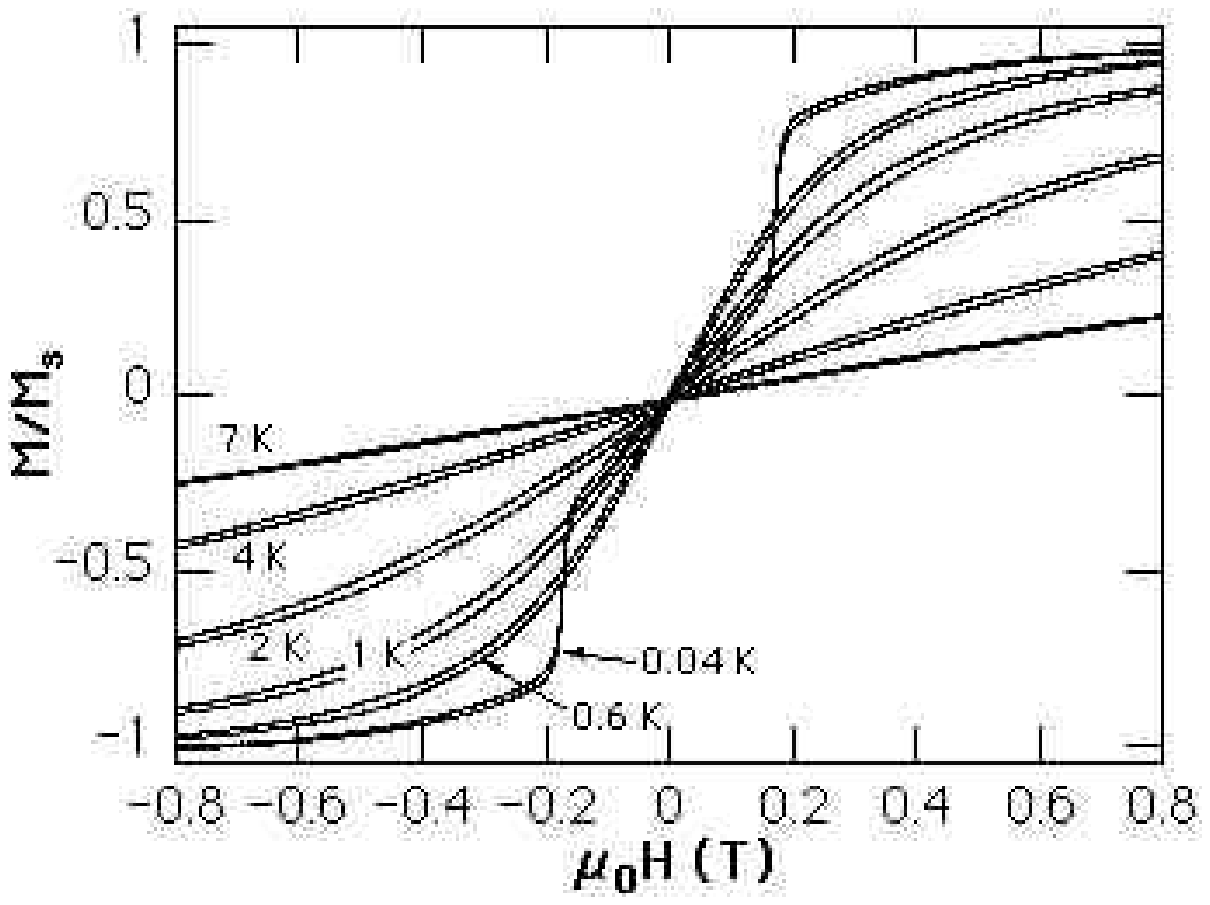}
\caption{Temperature dependence of the hysteresis loops, measured for two 
perpendicular directions in (100) plane. The temperatures are indicated.
In Figure 2a (above) the field is applied along the easy axis and the behaviour
is typical of a system with ferromagnetic intermolecular coupling. In Figure 
2b
(below), the field is applied perpendicular to the easy axis and the hysteresis
loops present an antiferromagnetic plateau.}
\label{hyst}
\end{center}
\end{figure}

\begin{figure}
\begin{center}
\includegraphics[width=.4\textwidth]{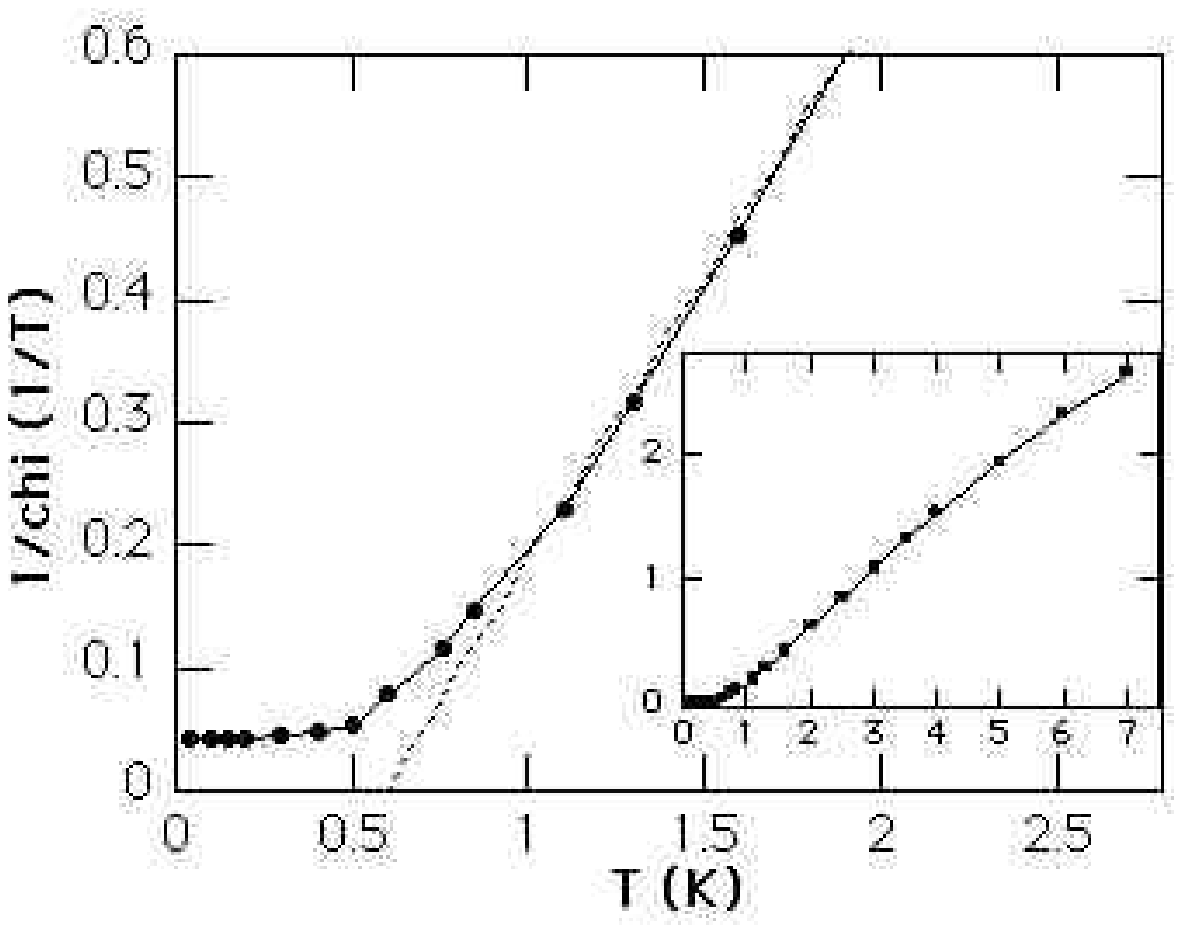}
\includegraphics[width=.4\textwidth]{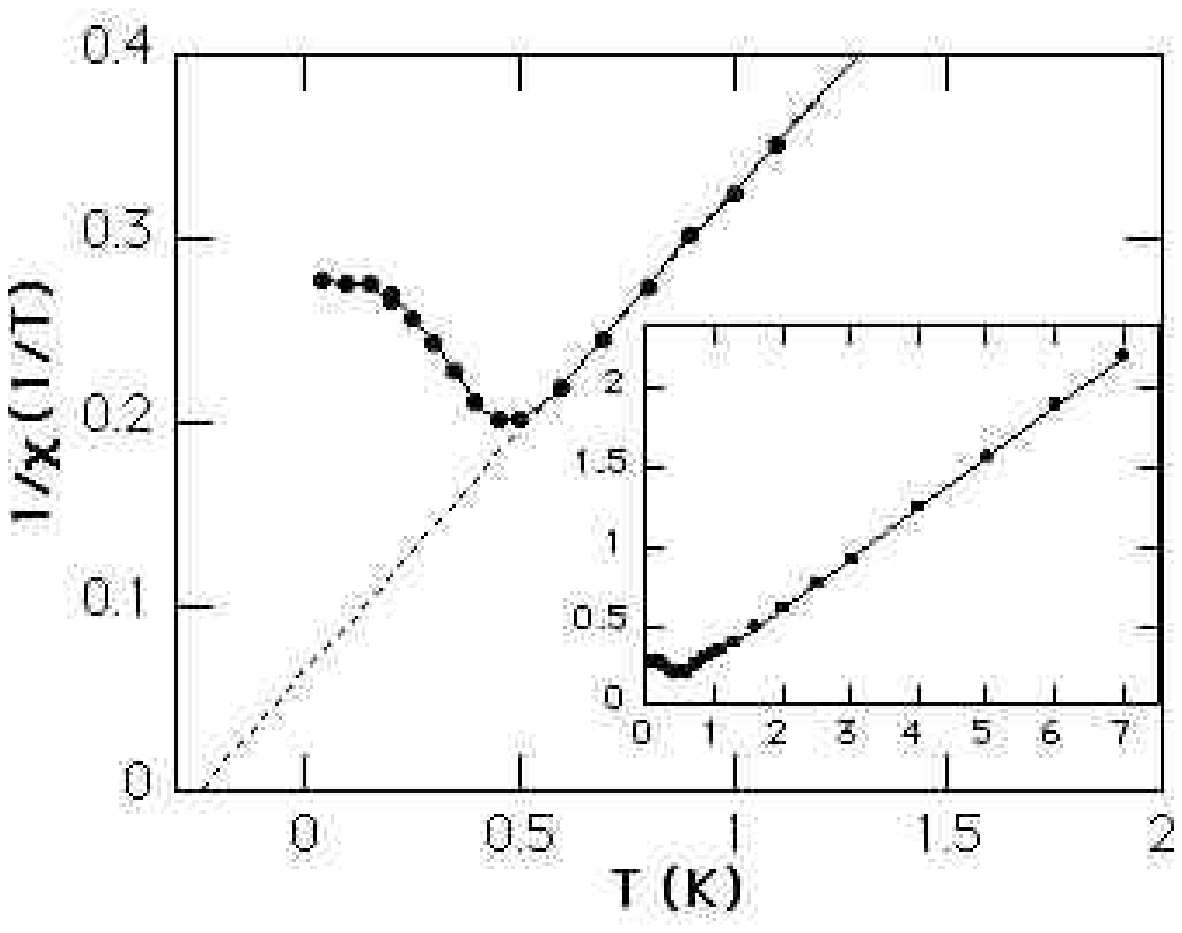}
\caption{Temperature dependence of the magnetic susceptibility in the two
orthogonal directions of Fig. 2. The dashed lines represent the fit with 
a Curie -Weiss law. Parallel to the easy axis (figure 3a, above), a positive 
Curie -Weiss constant is observed, (+ 0.6 K, ferromagnetic coupling). 
Perpendicular to the easy axis (figure 4b, below) the Curie -Weiss constant 
is negative (- 0.25 K, antiferromagnetic coupling).}
\label{fig 4}
\end{center}
\end{figure}

Three of the five complexes are involved in the present study:
i) [Cr(CN)$_4$[CN-Ni(tetren)]$_2$]Cl, noted CrNi$_2$(tetren) that crystallizes
in two different crystallographic systems: monoclinic, space group
= P2$_1$/c and orthorhombic, space group = Pbc2$_1$, named 1 and 1$^{*}$ respectively
; ii) [Cr(CN)$_4$[CN-Ni(dienpy$_2$)]$_2$](ClO$_4$) that crystallizes in the monoclinic 
system, space group = C2/c, noted CrNi$_2$(dienpy$_2$) or 3. This compound is
formed with a bulky ligand that influences the intermolecular distance and 
the intermolecular exchange coupling. The two cationic complexes are presented
in figure 1.
The CrNi$_2$(tetren) monoclinic system 1, forms pink distorted hexagonal crystals. 
The hexagonal face corresponds to the (1 0 0) crystallographic plane. The three 
others faces are the (-1-1 1), (1-1-1) and (1 1-1) planes. In the crystal, the
molecules are aligned in a two-dimensional sheet, parallel to the (1 0 0) plane.
The planes are separated by the counter anions, Cl$^{-}$ and water molecules that
form a network of hydrogen bonds. In each plane, 
the molecules are crystallographically equivalent but present two different 
orientations. The shorter intermolecular Cr-Cr, Cr-Ni and Ni-Ni distances are 
8.50, 6.23 and 7.50 \AA\ respectively. The shortest distance between two nearest
molecules is less that 3.08 \AA, corresponding to the distance between a free
cyanide ligand on the central chromium and a nitrogen atom of the tetren ligand.
Even if this distance appears slightly too long for a hydrogen bond, it 
can account for a possible pathway for intermolecular exchange interaction. 
For all the (1 0 0) planes, the direction of the molecular Ni-Cr-Ni axis 
forms a 36$^{\circ}$ angle with the a axis.\\
\indent The CrNi$_2$(tetren) orthorhombic system, 1$^*$, crystallises also as hexagonal plates.
The two dimensional molecular arrangement is exactly the same than in compound 1.
The shortest intermolecular Cr-Cr, Cr-Ni and Ni-Ni distances are 8.55, 6.29 and
7.50 \AA\ respectively. Only differs the crystallographic plane of the hexagonal 
face, which is (0 0 1) and the molecular orientation from one plane to another. 
Instead of having the same orientation, the molecules are aligned in chevrons, 
with an angle of 90$^{\circ}$ between the molecular Ni-Cr-Ni axes. As in the previous 
case, the chloride ions and the hydrogen bond network formed by the water
molecules separate the molecular layers.

Finally, the crystals obtain from the CrNi$_2$(dienpy$_2$) system, 3, 
are red-pink parallelepipeds that belong to the monoclinic system. 
The molecules are aligned in a two dimensional
arrangement and the planes are separated by the ClO$_4$-counterions. 
The orientations of the molecules in the crystal are very similar to the one
observed in compound 1. Despite the facts that the dienpy$_2$ ligand, due to 
the presence of pyridine rings, is more bulky than the tetren ligand and 
that the perchlorate is bigger than chloride, the intermolecular distances
in 3 are only slightly longer than in complexes 1 and 1$^{*}$. The shorter 
intermolecular Cr-Cr, Cr-Ni and Ni-Ni distances are 9.17, 6.84 and 8.56 \AA\
respectively. This is due to p-p interaction and to hydrogen bonds between
two free cyanides through a water molecule.\\

{\bf 3.  CrNi$_2$ complexes, magnetic behaviour}

\indent The magnetisation versus applied field was recorded on single crystals of 
the three CrNi$_2$ using an array of micro-SQUIDs ~\cite{31}. Different from a 
classical SQUID magnetometer where the flux coupling between sample and
SQUID is made by using a pick up coil, a micro-SQUID magnetometer allows 
a much better flux coupling since the sample is placed directly on the 
SQUID loops. The sensitivity achieved with our device is 10$^{-17}$ emu, that 
is ten orders of magnitude better than a traditional SQUID magnetometer 
~\cite{32} The sensitivity is smaller when the sample is much larger than the 
micro-SQUID. The high sensitivity of the magnetometer allows to study 
single crystals with dimensions from 10 to 500 $\mu$m. The magnetometer works
in a temperature range between 0.035 and 10 K, in applied fields up to
1.4 T with sweeping rates as high as 1 T s$^{-1}$ and a field stability better
than a microtesla. The field can be applied in any direction of the 
micro-SQUID plane by separately driving three orthogonal coils. In 
order to ensure a good thermalisation, the crystal is fixed by using 
Apiezon grease.

\begin{figure}
\begin{center}
\includegraphics[width=.4\textwidth]{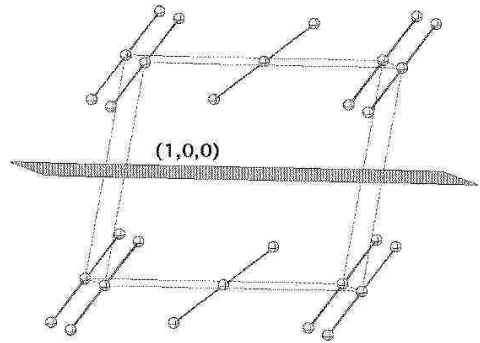}
\caption{Schematic perspective view of compound 1. The molecules are represented
by a straight-line nickel-chromium-nickel. Ligands are removed for clarity. 
There are two different orientations of the molecules. The angle between the
two directions is 64$^{\circ}$; the angle between each direction and the (100) plane 
is 36$^{\circ}$.}
\label{fig 5}
\end{center}
\end{figure}

Here we report on the observation of canted ferromagnetic structure for the 
three different compounds of CrNi$_2$ with the same ground state spin 7/2. First
we analyse the hysteresis loops of compound 1. Second, we correlate the 
structure with the magnetic behaviour. Third, we conclude with a comparison 
between the magnetisation data of the three crystals in order to establish 
if the intermolecular coupling is sensisitive to small variations of the 
structure: terminal ligands or crystal symmetry.

We used the micro-SQUID to measure magnetisation hysteresis loops at 
different temperatures and different field sweep rates and also the magnetic
susceptibility below 7 K. A single crystal was placed on the SQUID array
with the (100) plane parallel to the SQUID plane. The hysteresis loops
are independent on the field sweep rate, suggesting that either resonant
quantum tunnelling is hindered by intermolecular interactions or is much 
faster than the time scale of the micro-SQUID technique.

The magnetisation loops measured for different directions of the applied 
field in the (100) plane, proved that the magnetic behaviour is not isotropic
in the plane. We have chosen two symmetry directions in the (100) plane which
present distinct magnetic behaviour. The first direction is along the projection
in the (100) plane of the bisecting line of two Ni - Cr - Ni molecular directions,
showed to be an easy axis of magnetization. The second direction perpendicular
to the first one, showed to be a hard axis of magnetization. The corresponding
magnetizations versus applied field curves are presented in figure 2(a, b). 
In the first orientation (figure 2a), the magnetization is reversed from 
the negative saturation value to the positive saturation value in zero field. 
When increasing the temperature, the saturation magnetization decreases,
but the zero field slope remains constant below 0.6 K. This is confirmed
by the magnetic susceptibility measured in this direction (see figure 
3a).

\begin{figure}
\begin{center}
\includegraphics[width=.4\textwidth]{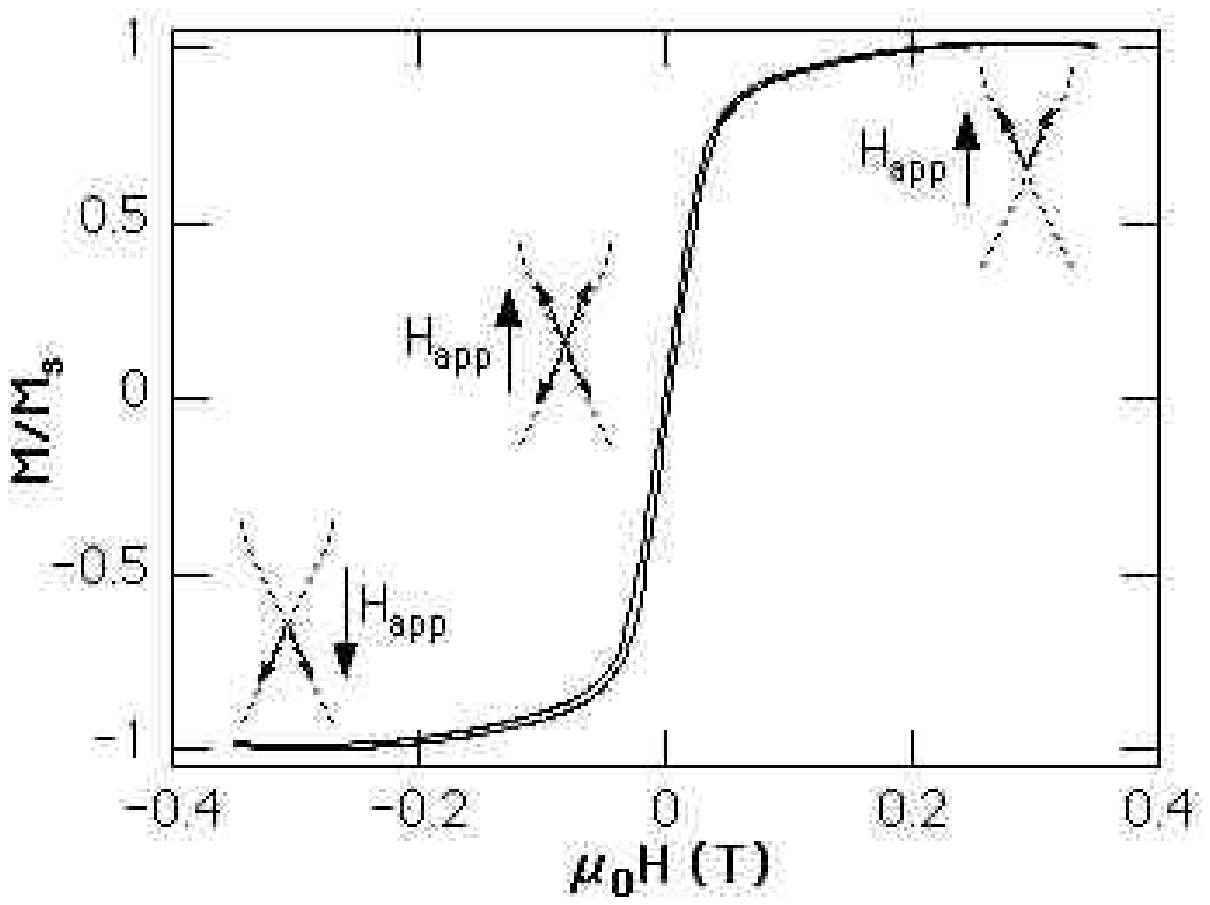}
\includegraphics[width=.4\textwidth]{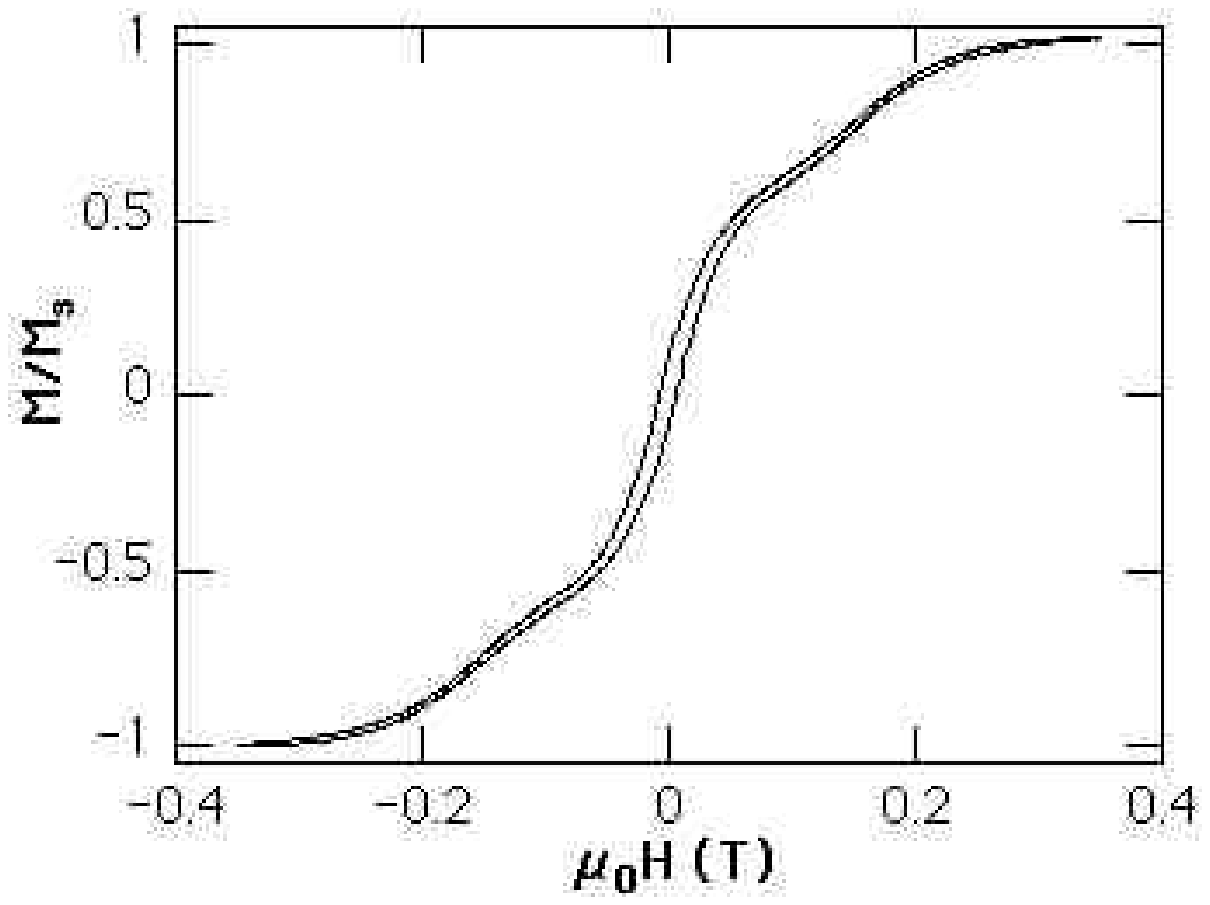}
\includegraphics[width=.4\textwidth]{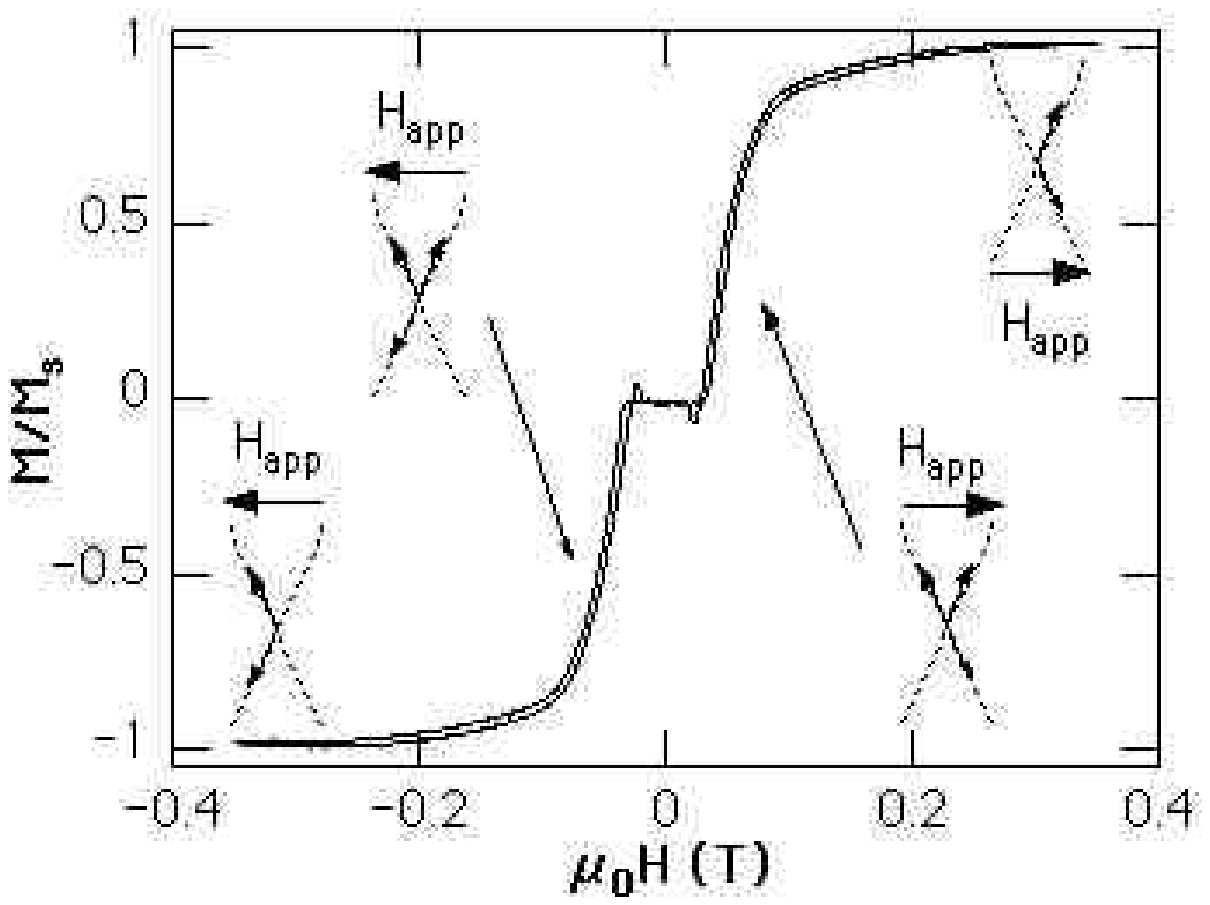}
\caption{Hysteresis loops for different orientations of the field : 
parallel to a) the easy axis; b) intermediate axis; 
c) hard axis. The insert depicts the
orientations of the molecular moments relative to the applied field Happ. 
The easy axis corresponds to the bisecting line of two in plane moments 
(fig. 5a). Perpendicular to this direction, only the net antiferromagnetic 
component of the coupling is sensitive to the field variation (fig. 5c).
For intermediate orientations (fig. 5b), the behaviour is a mixture of the 
two limiting cases.}
\label{fig 6}
\end{center}
\end{figure}

A fit to a Curie-Weiss law gives a positive Curie-Weiss constant of +0.6 K.
Both results are typical of a system with ferromagnetic intermolecular 
coupling leading to long range ferromagnetic ordering. However, when 
the field is applied perpendicular to the easy axis of magnetization 
(second direction above), hysteresis loops (figure 2b) and susceptibility 
measurements (figure 3b) suggest an antiferromagnetic intermolecular 
coupling. This is established by the hysteresis loop shapes and the negative
Curie-Weiss constant (-0.25 K).

The magnetic behaviour can be explained using the crystallographic 
structure. The CrNi$_2$ compound, 1, presents an uniaxal anisotropy with a small
ferromagnetic exchange interaction between the molecules. The central chromium
ions of the molecules define the (100) planes (figure 1), which are separated
from each other by a water and halide layer (figure 4). There might be weak 
antiferromagnetic interactions via hydrogen bonds between the planes but that
are not clearly established. There are two different orientations of the
molecules centred in the same (100) plane. The angle between the two directions 
is 64$^{\circ}$; the angle between each direction and (100) plane is 36$^{\circ}$.

The magnetic moments have therefore two orientations corresponding to the two 
orientations of the molecules. The micro-SQUID magnetometer is sensitive 
only to the in plane component of the magnetization, in this case to the 
projected magnetisation into the (100) plane. The projection coupling 
of magnetic moments can be decomposed into two independent contributions:
a net ferromagnetic coupling along the bisecting line of two moments and 
a net antiferromagnetic one perpendicular to this direction. The magnetization
measured along the bisecting line of two moments (figure 5a) is sensitive
only to the net ferromagnetic component (the antiferromagnetic one is
perpendicular to the direction of magnetic field, that is insensitive 
to field variation). The two moments are reversed at the same time near 
zero field.

\begin{figure}
\begin{center}
\includegraphics[width=.4\textwidth]{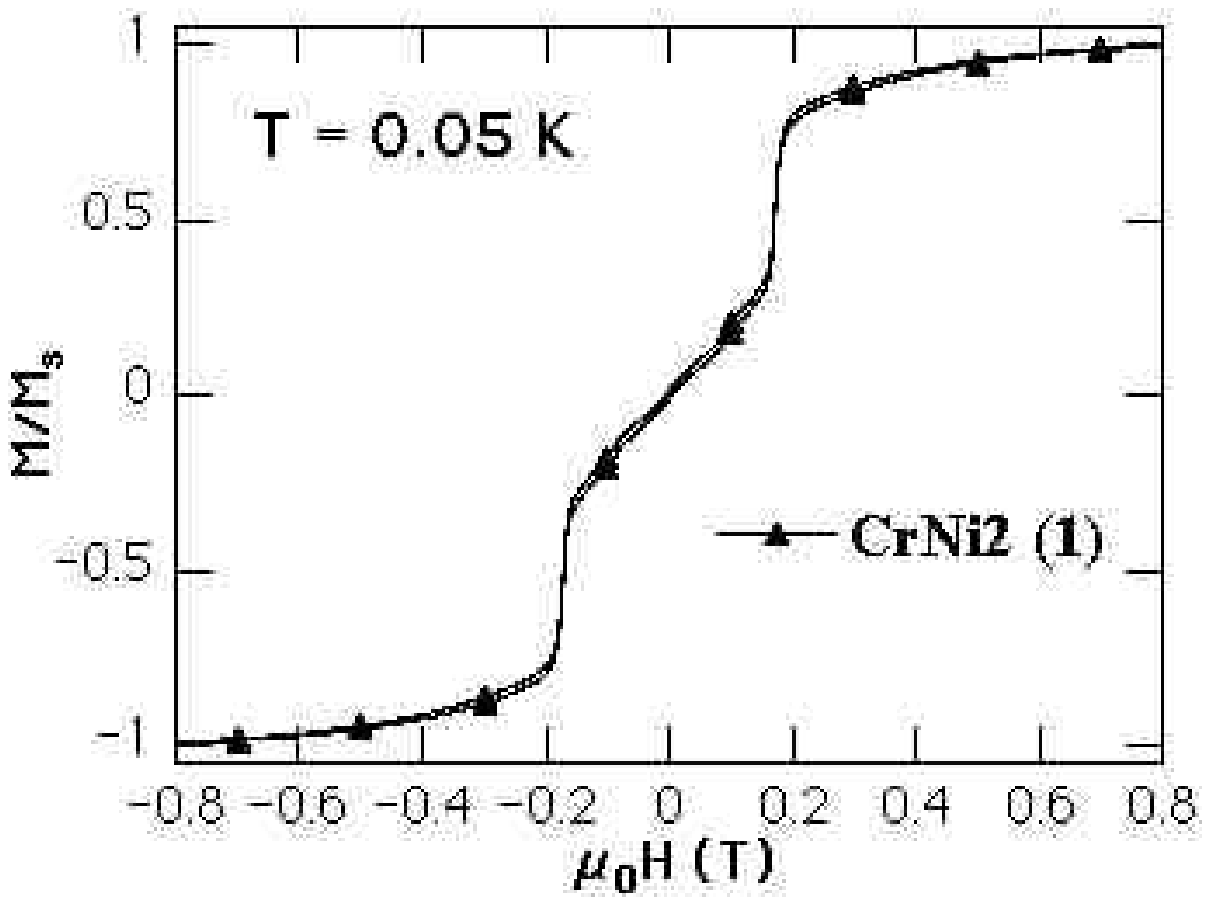}
\includegraphics[width=.4\textwidth]{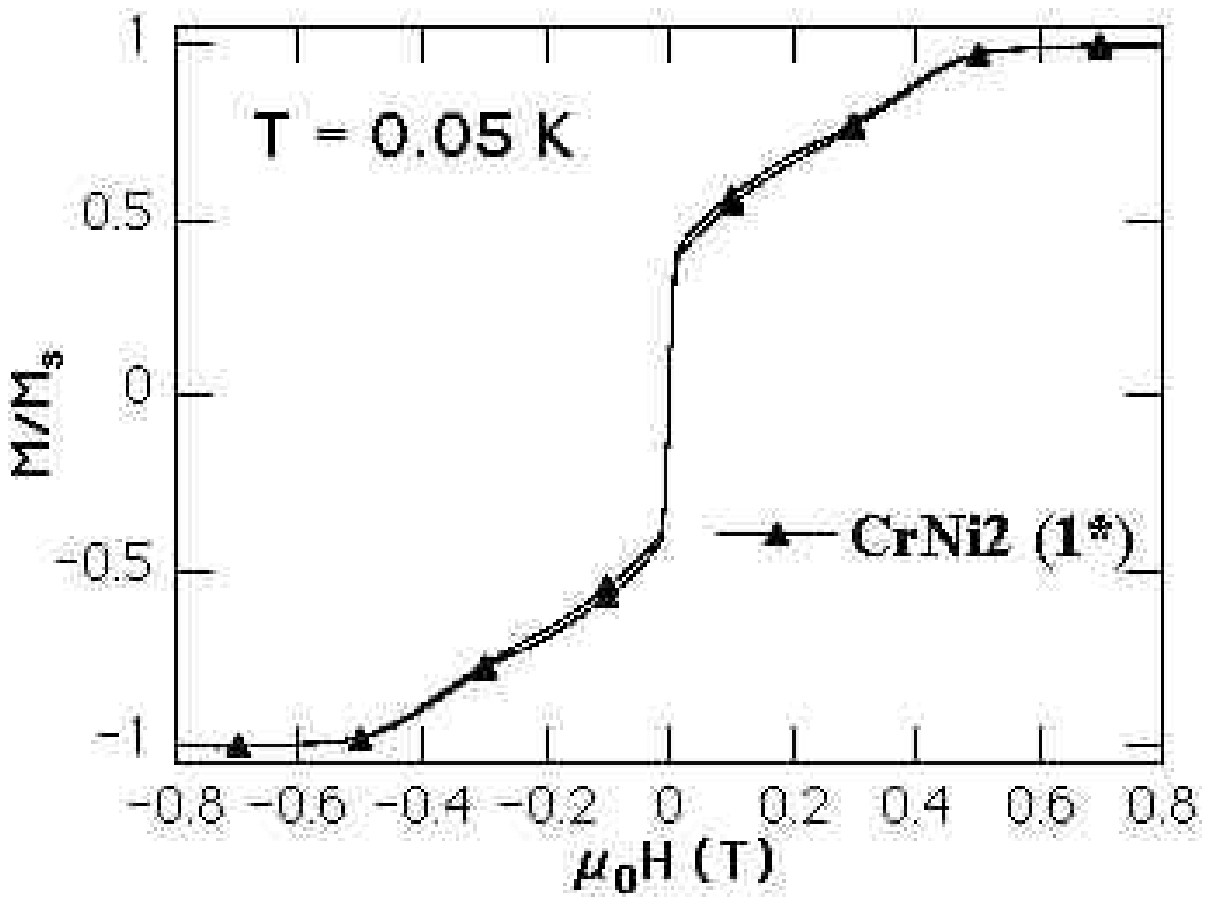}
\includegraphics[width=.4\textwidth]{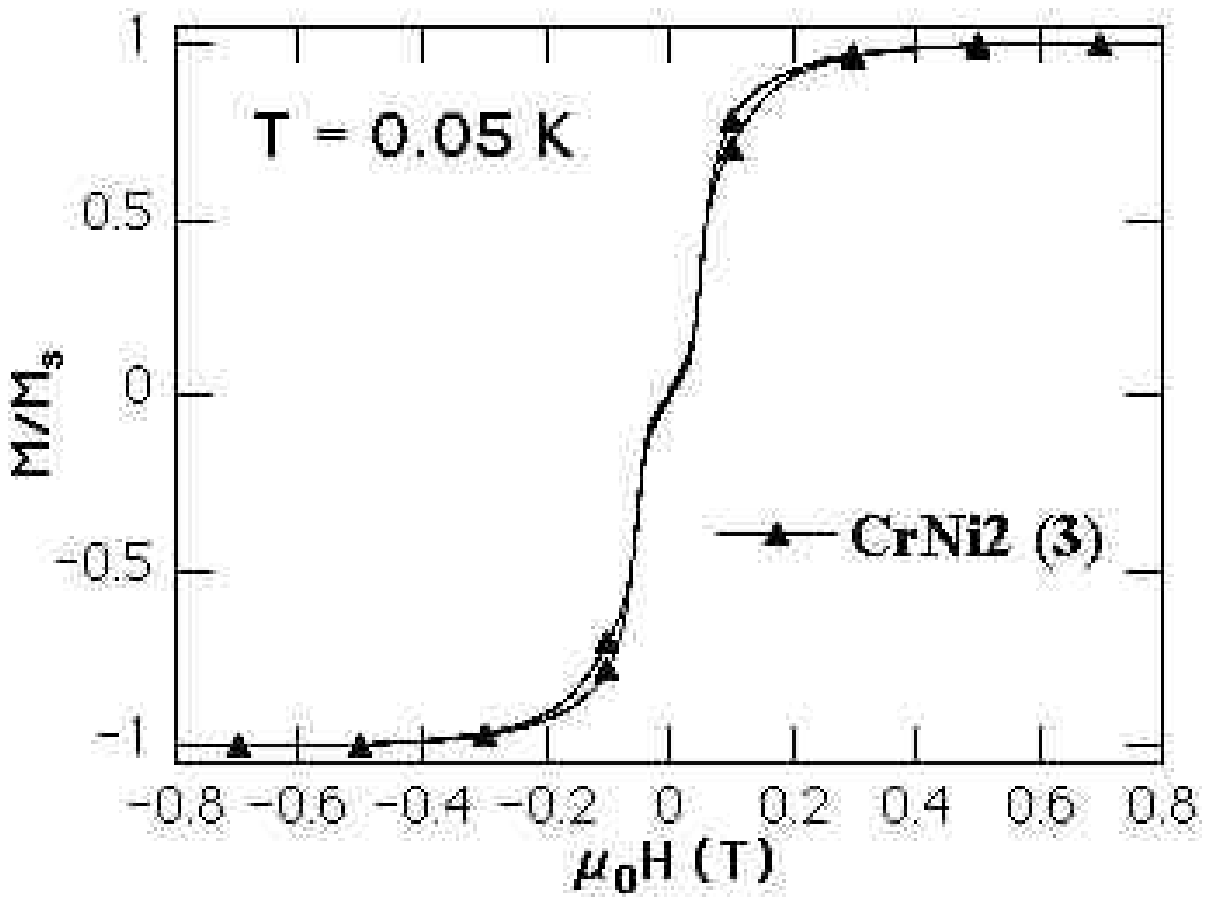}
\caption{Hysteresis loops for the three samples 1, 1$^*$ and 3; the magnetic
field is applied along the hard axis. The field amplitude of the antiferromagnetic plateau
for compound 1$^*$ is lower than for compound 1 and higher than for compound 3 
in agreement with the estimation of exchange interaction coupling.}
\label{fig 7}
\end{center}
\end{figure}

Perpendicular to the bisecting line (figure 5c) only the net antiferromagnetic
component of the couple is sensitive to the magnetic field. When the field
increases (respectively decreases), one of the spins flips in order to align
the couple with the maximum of the ferromagnetic component parallel to the 
field.
For any other orientation, the behaviour is a mixture of the two limiting cases.
At zero field we recover the projection of the ferromagnetic component of
the couple on the field direction. When the field increases one of magnetic 
moment flips and the couple changes; the ferromagnetic projections on the 
field direction increase.

Measured in similar conditions, compounds 1$^{*}$ and 3 shows the same behaviour 
(figure 6). The field amplitude of the antiferromagnetic plateau for compound
1$^{*}$ is lower than for compound 1 and higher than for compound 3 in good 
agreement with the estimation of exchange interaction constant and with the fact
that the molecules are closer in compound 1 than in compound 3. This points
out the necessity to use bulkier ligands to reduce the intermolecular 
interaction.
Other studies are in progress to better understand the correlation between
the crystallographic structure and the magnetic behaviour in other
directions of space.\\

{\bf 4. Conclusions}

\indent The two different orientations of molecules in the crystals together
with the intermolecular exchange interactions make CrNi$_2$ a model system 
for canted ferromagnetic structures. It is established that the intermolecular
coupling is sensitive to small variations of the structure such as the 
crystal symmetry or the steric hindrance of the terminal ligand. Our study
allows to better understanding the synthetic parameters that have to be tuned 
in order to minimise intermolecular exchange interactions, necessary condition
to get single molecule magnets. The ligand, as well as the counter anions has
to be as bulky as possible. To isolate the high spin molecules, it is also
conceivable to dilute the molecules in a diamagnetic or a paramagnetic matrix.
Preliminary results performed on the CrNi$_2$(tetren) compound co-crystallised 
in presence of CoNi$_2$(tetren) complex, viewed as an ordered paramagnetic medium
indicate the validity of such a strategy, which is beyond the scope of 
the present communication.

\end{document}